\documentclass[12pt]{iopart}
\usepackage{iopams}
\usepackage{graphicx, caption}
\usepackage{subcaption}
\usepackage{float}
\usepackage{lipsum} 
\usepackage[normalem]{ulem}
\usepackage{xcolor}
\newcommand{\beq}{\begin{eqnarray}}
\newcommand{\eeq}{\end{eqnarray}}
\newcommand{\be}{\begin{eqnarray*}}
\newcommand{\ee}{\end{eqnarray*}}
\newcommand{\bal}{\begin{align}}
\newcommand{\eal}{\end{align}}

\begin{document}



\title[Physica Scripta 2023]{Correlation between the nuclear structure and reaction dynamics of Ar-isotopes as projectile using the relativistic mean-field approach}
\author{ }
\date{\today}
\author{Monalisa Das$^{1}$, J. T. Majekodunmi$^2$, N. Biswal$^{1}$, R. N. Panda$^{1}$, and M. Bhuyan$^{3}$  
}
\noindent
\address{$^1$Department of Physics, Siksha $'O'$ Anusandhan, Deemed to be University, Bhubaneswar-751030, India}
\address{$^2$Institute of Engineering Mathematics, Faculty of Applied and Human Sciences, Universiti Malaysia Perlis, Arau, 02600, Perlis, Malaysia}
\address{$^3$Center of theoretical and Computational Physics, Department of Physics, University of Malaya, Kuala Lumpur, 50603, Malaysia}

\ead{bunuphy@um.edu.my}
\vspace{10pt}

\begin{abstract}
This theoretical study is devoted to bridging the gap between the nuclear structure and reaction dynamics and unravelling their impact on each other, considering the neutron-rich light mass $^{30-60}$Ar isotopes. Using the relativistic mean-field with the NL3$^*$ parameter set, several bulk properties such as binding energies, charge radii, quadrupole deformation parameter, two neutron separation energy, and differential two neutron separation energy with the shell closure parameter are probed for the mentioned isotopic chain. For validation, the RMF (NL3$^*$) results are compared with those obtained from the finite range droplet model (FRDM), Weizs$\ddot{a}$cker-Skyrme model with WS3, WS$^*$ parameters and the available experimental data. Most of the participating isotopes are found to be prolate in structure and neutron shell closures are conspicuously revealed at $N=14, 20, 40$ but weakly shown at $N=24, 28, 34$.  From our analysis, a central depletion in the nucleonic density is identified in $^{32}$Ar and $^{42-58}$Ar, indicating them as possible candidates for a semi-bubble-like structure. Interestingly, these results are consistent with recent theoretical and experimentally measured data. Besides, using the Glauber model, the reaction cross-sections are determined by taking $^{26-48}$Ar as projectiles and stable targets such as $^{12}$C, $^{16}$O, $^{40}$Ca, $^{90}$Zr, $^{124,132}$Sn, $^{208}$Pb and $^{304}$120. Although there is no experimental evidence for the stability of  $^{304}$120, it has been predicted in Ref. [Mod. Phys. Lett. A {\bf 27},  1250173  (2012)] as a stable nucleus. A relatively higher cross-section value is noticed between $^{30}$Ar and $^{32}$Ar which infers that $^{32}$Ar is the most stable isotope among the considered chain. Moreover, we noticed that the profile of the differential cross-sections and scattering angle are highly influenced by the mass of the target nuclei and the magnitude of the incident energy of the projectile nucleus.
\end{abstract}

\section{Introduction}
One of the intriguing goals of nuclear physics is to seek a conceptual understanding and unravel the complexities associated with the many-body systems. The atomic nucleus is composed of quantum constituents (nucleons) that manifest various characteristics (such as shape coexistence, bubble structure e.t.c). In particular, the configuration of nucleons within the nucleons of these systems, (especially those lying close to the drip-line (exotic or halo nuclei)) opens an avalanche of insight into the nuclear structure as well as their reactions \cite{holt12}.  The production of these exotic nuclei has been aided by the advancement of radioactive ion beam (RIB) facilities sequel to its first detection \cite{wils46}. The observation of bubble structure is often characteristic of atomic nuclei with unconventional depletion of the central densities which can be largely attributed to their unoccupied $s-$orbital near the Fermi surface \cite{saxe19}.  The density at the central region assumes a relatively lower magnitude than the saturation density thereby forming a dip/hollow structure inside the nucleus. A plausible explanation for the depopulated $s-$orbit could be linked with the inversion $s_{1/2}$ with other states located above it \cite{khan08}. This presupposes that the $s-$orbit near the Fermi energy must be encompassed with orbitals having relatively larger $\ell$ (angular momentum value) such that the dynamic correlations are weakened. Besides, several investigations on heavy and superheavy nuclei reveal the existence of bubble-like structure results from the interaction between the Coulomb and nucleon-nucleon (NN-) interaction \cite{sobi07,dech99,sing12,ikra14,bend13}. However, this phenomenon cuts across the entire nuclear territory including the light, heavy and superheavy nuclei.

On the other hand, the pairing correlation and deformation effects are known to retard bubble formation \cite{saxe19}. The emergence of bubble structure is usually quantified in terms of the depletion fraction (D.F) and the charge density distribution which are susceptible to the quantal effects arising from the filling of single-particle levels near the Fermi energy. Moreover,  one of the notable features of exotic nuclei is the imbalance between the ratio of the proton  and neutron number. Consequently, in the exotic light nuclei, such structural changes can result in a quenching or total disappearance of the shell closures and  new magic numbers may evolve. The literature is replete with experimental and theoretical shreds of evidence showing that some exotic nuclei like $^{24}$O \cite{hoff09,kanu09}, $^{42}$Si \cite{camp06,take12,stro14,bast07}, $^{44}$S \cite{forc10,kimu13,sara00,gaud09} do not reveal the usual shell structure behaviour of the traditional magic numbers. Following these conjectures, the Argon (Ar) isotopes are considered as  suitable candidates for the bubble structure phenomenon in the present study. Thus, a concerted effort is given to establish a bridge/correlation between the nuclear structure and reaction dynamics.

Furthermore, noble gases such as neon, argon and xenon have a perfect impedance for weakly interacting massive particle targets. Klein {\it et al.} \cite{klei96} was given a remarkable effort to measure the isotopic shift and hyper-fine structure of $^{32-40,46}$Ar. The authors reported that the variation in the mean square nuclear charge radii from N= 14-22 and 28 changes in the magnetic dipole and electric quadrupole moments for odd-argon-isotopes. Recent investigations \cite{saxe19,thak19,thak20,adri20} of $^{38,39,40-44}$Ar and $^{46}$Ar-isotopes include the entire $f_{7/2}$ shell with an anomaly at $^{45}$Ar-isotope. The nuclei such as $^{36,38,40}$Ar do not exhibit the nuclear quadrupole moments whereas $^{37,39}$Ar-isotopes are radioactive having half-life values 296 y, 35.0 d and their spins are 7/2 and 3/2 respectively. It has been demonstrated \cite{mraz04} that despite the weak energies in the lowest excited state of $^{45}$Ar, there is an appreciable amount of energy that demands due consideration. Thus, it is necessary to extend the previous works \cite{thak19,thak20,adri20}  to explore the structural and reaction mechanism of Ar-isotopes. The theories adopted for nuclear structure calculations can be categorised into two groups. Firstly, the {\it ab-initio} methods \cite{pudl95,pudl97} which include the QMC methods, coupled cluster methods and no-core shell model \cite{carl15,hage16,barr13}. On the other hand, those emanating from the energy density functional approach include the relativistic \cite{bend03,sero92,ring96,posc97} and the non-relativistic mean field theories \cite{doba84,davi85,nege82}. Among these existing nuclear structure models, we have employed the relativistic mean-field (RMF) formalism which self-consistently incorporates the spin-orbit term and the Bardeen-Cooper-Schrieffer (BCS) \cite{ring96,saxe17} theory with the NL3$^*$ parameter set \cite{lala09}. Since it has been found effective for the treatment of ground state nuclei lying close to the driplines and provides an accurate description of the nuclear many-body dynamics across the entire nuclear landscape \cite{ring96,josh22,maje22}.

Besides, the study of nuclear reaction cross-sections, including the total reaction, differential elastic scattering and Coulomb break-up cross-sections have been studied by the secondary radioactive beam technique. A detailed analysis of these quantities can provide rich insight into the structural properties of unstable nuclei, especially regarding those with halo shapes nearer to the drip-line  \cite{tani96,hans89,furn97,arum04}. The theoretical frameworks developed for the description of the reaction of halo nuclei vary with the projectile incident energy as well as the reaction mechanisms \cite{ogaw01}. According to the research of Ogawa et al. \cite{ogaw01} the Coulomb breakup process and other low-energy reactions are described using the time-dependent Schrodinger equation with and without perturbation approximation, while high-energy reactions, such as those involving nuclear fusion are described using the Glauber multiple scattering theory \cite{glau59}. It is worth noting that the Glauber multiple scattering theory provides an adequate footing for the microscopic description of the reaction. The neutron halos can be formed by using weakly bound neutrons which are dissociated from the nuclear core.  Previous analysis \cite{tani96a,tani92} reveals that the interaction cross-section for $^{6,8}$He, $^{11}$Li, $^{11,14}$Be have greater values  due to their larger values of rms-radii. Similarly, $^{19}$C, $^{22}$N, $^{23}$O, $^{24}$F are characterized by high matter radii and interaction cross-sections signifying their one neutron halo feature \cite{ozaw01}. The $^{22}$C having a magic neutron number (N=16) is a newly formed halo nucleus with a configuration like the Borromean halo structure whereas $^{21}$C is an unstable nucleus \cite{ozaw00,tani01}. The bulk properties of nuclei such as the binding energy, root mean square charge radii, quadrupole deformation, two neutron separation energy, differential two neutron separation energy, shell closure parameter, depletion fraction, and the reaction cross-section and differential cross-sections are compared with their corresponding experimental data wherever available. Here, the RMF(NL3$^*$) formalism is used for the structural analysis, and further, the structural input is used within the Glauber model to analyse the scattering phenomenon for the Ar-isotopes as a projectile. \\
This paper is designed as follows:  Sec. \ref{theory}, briefly presents the Relativistic mean-field and Glauber Model. The result from our calculations is presented and discussed in Sec. \ref{result}. Finally, Sec. \ref{summary} gives an overview/summary of this study.
\section{Relativistic mean field model}
\label{theory} 
For the nuclear structure analysis, the relativistic mean-field (RMF) model plays a crucial role as explained in the preceding section. The interaction between the nucleons and mesons can be described by the non-linear Lagrangian density \cite{bhuy11,bhuy15,bhuy18,pann87,lala99c,bhuy18a,rein89,vret05,meng06,paar07,niks11,logo12,zhao12,bogu77,sero86,carl00,patr09,ring96,josh22,maje22,lala09}, 
\begin{eqnarray}
{\cal L}&=& \overline{\psi}\{i\gamma^{\mu}\partial_{\mu}-M\}\psi +{\frac12}\partial^{\mu}\sigma
\partial_{\mu}\sigma \nonumber \\
&& -{\frac12}m_{\sigma}^{2}\sigma^{2}-{\frac13}g_{2}\sigma^{3} -{\frac14}g_{3}\sigma^{4}
-g_{s}\overline{\psi}\psi\sigma \nonumber \\
&& -{\frac14}\Omega^{\mu\nu}\Omega_{\mu\nu}+{\frac12}m_{w}^{2}\omega^{\mu}\omega_{\mu}
-g_{w}\overline\psi\gamma^{\mu}\psi\omega_{\mu} \nonumber \\
&&-{\frac14}\vec{B}^{\mu\nu}.\vec{B}_{\mu\nu}+\frac{1}{2}m_{\rho}^2
\vec{\rho}^{\mu}.\vec{\rho}_{\mu} -g_{\rho}\overline{\psi}\gamma^{\mu}
\vec{\tau}\psi\cdot\vec{\rho}^{\mu}\nonumber \\
&&-{\frac14}F^{\mu\nu}F_{\mu\nu}-e\overline{\psi} \gamma^{\mu}
\frac{\left(1-\tau_{3}\right)}{2}\psi A_{\mu},
\label{lag}
\end{eqnarray}
with vector field tensors
\begin{eqnarray}
F^{\mu\nu} = \partial_{\mu} A_{\nu} - \partial_{\nu} A_{\mu} \nonumber \\
\Omega_{\mu\nu} = \partial_{\mu} \omega_{\nu} - \partial_{\nu} \omega_{\mu} \nonumber \\
\vec{B}^{\mu\nu} = \partial_{\mu} \vec{\rho}_{\nu} - \partial_{\nu} \vec{\rho}_{\mu}.
\end{eqnarray}
Here the $\sigma$, $\omega_{\mu}$ and $\vec{\rho}_{\mu}$ are the fields for the $\sigma$-, $\omega$- and isovector $\rho$- meson respectively, with $A_{\mu}$ is the electromagnetic field. The corresponding field tensors for the $\omega^{\mu}$, $\vec{\rho}_{\mu}$ and photon fields are $\Omega^{\mu\nu}$, $\vec{B}_{\mu\nu}$ and $F^{\mu\nu}$. These sets of equations are solved self-consistently. The centre-of-mass motion energy $E_{c.m.}$ correction is calculated by using the standard harmonic oscillator formula i. e $E_{c.m.}=\frac{3}{4}(41A^{-1/3})$. The quadrupole deformation parameter $\beta_2$ is evaluated as a combined effect of proton and neutron quadrupole moments $Q = Q_n + Q_p =\sqrt{\frac{16\pi}5} (\frac3{4\pi} AR^2\beta_2)$ and the root mean square (rms) matter radius, $\langle r_m^2\rangle = {1\over{A}}\int\rho(r_{\perp},z) r^2d\tau$. Here $A$ represents the mass number, and  $\rho(r_{\perp},z)$ is the axially deformed density. We reiterate that the NL3$^*$ \cite{lala09} parameter is used for the present calculation. This parameter set adequately describes the ground state properties of the nuclei from $\beta-$ stable region to the drip-line \cite{pann87,lala99c,patr09,bhuy11,bhuy15,bhuy18,bhuy18a,rein89,ring96,vret05,meng06,paar07,niks11,logo12,zhao12,lala09}.  

\subsection{Glauber Model}
The Glauber theory \cite{glau59} was proposed to describe the quantum theory of collisions of composite particles. This theory provides a systematic analytical expression that considers the many-body nuclear system either as a projectile or target and is efficient for studying heavy ion elastic scattering and reaction cross-section ($\sigma_{R}$). Particularly, it is used to determine the momentum of nuclei and the direction of collisions. The generalised formula for $\sigma_{R}$ is given as,
\begin{equation}
\sigma_R=2\pi\int_0^\infty \textbf{b}[1-T(\textbf{b})]d\textbf{b}, \label{gm}
\end{equation}
where T(\textbf{b}) is the transparency function and \textbf{b} refers to the impact parameter. Although the Glauber model was initially proposed for the description of collisions of halo nuclei at high incident energy range yet, its application spans across different energy ranges due to Coulomb corrections \cite{chau83,buen84}. From eq. (\ref{gm}), the T(\textbf{b}) is  expressed as,  
\begin{equation}
T(\textbf{b})=|e^{\iota \chi_{PT}(\textbf{b})}|^2.
\label{trns}
\end{equation}
Here, $\chi_{PT}$ symbolizes the projectile-target phase shift function  and takes the expression 
\begin{equation}
\iota\chi_{PT}(\textbf{b})=-\sum_{i,j}\sigma_{NN}\int\overline{\rho_{P}}(\textbf{s})
\overline{\rho_{T}}(|\textbf{b}-\textbf{s}|) d\textbf{s}.
\end{equation}
The summation includes nucleons i.e $i$ and $j$ refers to projectile and target nuclei respectively. More details can be found in our previous works \cite{pand14,shar16,patr09}.

The differential elastic scattering cross-section is presented about Rutherford's scattering cross-section as,
\begin{eqnarray}
\frac{d\sigma }{d\Omega }=\frac{\left| F({\bf q})\right| ^{2}}{\left|
F_{coul}({\bf q})\right| ^{2}} \;.
\label{diff}
\end{eqnarray}
Here, the variable $F(q)$ and $F_{coul}({\bf q})$ denotes the elastic and Coulomb (elastic) scattering amplitudes respectively. \textbf{q} is the momentum transferred from the projectile to the target. With the inclusion of the Coulomb interaction, the elastic scattering amplitude  $F({\bf q})$ is of the form,
\begin{eqnarray}
F({\bf q})=e^{i\chi _{s}}\left\{ F_{coul}({\bf q}) 
+\frac{iK}{2\pi }\int db e^{-iq.b+2i\eta {\ ln (Kb)}}T(b)\right\},
\end{eqnarray}
and further simplification of the Coulomb elastic scattering amplitude $F_{coul}({\bf q})$ yields
\begin{eqnarray}
F_{coul}({\bf q})=\frac{-2\eta K}{q^{2}}\exp \left\{ -2i\eta {\ ln}\left( \frac{q}{%
2K}\right) +2i\arg \Gamma \left( 1+i\eta \right) \right\}\;.
\end{eqnarray}
Here, $\eta $=$Z_{P}Z_{T}e^{2}$/$\hbar$ is the Sommerfield parameter and $v$ typifies the incident velocity of the projectile. Here $\chi _{s}=-2\eta \ln (2Ka)$ and $a$ being the screening radius \cite{glau59}, whereas, K is the momentum of the projectile. The differential elastic cross section is independent of the screening radius $a$.

\begin{table*}
\caption{The binding energy, charge radii, quadrupole deformation parameter ($\beta_{2}$), $\Delta BE$ = BE$_{Exp}$ - BE$_{Th}$ and the depletion parameter for the isotopic chain of Ar-nucleus are noted within the relativistic mean-field approach for the NL3$^*$ parameter set. The theoretical predictions from FRDM \cite{moll15}, WS mass formula for WS3 \cite{liu11}, WS$^*$ \cite{wang10} parameters and the available Expt. data \cite{wang12,ange13,rama01} are listed for comparison.}
\renewcommand{\tabcolsep}{0.03cm}
\renewcommand{\arraystretch}{1}
\begin{tabular}{cccccccccccccc}
\hline
Nuclei&\multicolumn{4}{c}{Binding energy (B.E.)} &\multicolumn{3}{c}{Charge Radii (R$_{ch}$)} &\multicolumn{4}{c}{Quadrupole Deformation ($\beta_2$}) & \multicolumn{1}{c}{$\Delta BE$}
&\multicolumn{1}{c}{$(D.F)_T$} \\ 
&RMF&FRDM&WS3 &Expt.&RMF&WS$^*$ &Expt.  &RMF&FRDM&WS3&Expt. & & RMF \\
& (NL3$^*$)& \cite{moll15}&\cite{liu11}&\cite{wang12}&(NL3$^*$) &\cite{wang10}&\cite{ange13}& (NL3$^*$)&\cite{moll15}&\cite{liu11} &\cite{rama01} & &(NL3$^*$) \\
\hline
$^{30}$Ar &206.41 &207.99&204.83 &206.58& 3.711&3.391& & 0.257&-0.280 &0.159  &  &  0.18 &-- \\
$^{32}$Ar & 243.68 & 246.12&245.42 &246.40& 3.639&3.353&3.347 &0.051&-0.281& -0.087 & & 2.722 & 0.048 \\
$^{34}$Ar & 274.87 &278.25&278.71& 278.72& 3.601&3.373& 3.365&0.027&-0.230&  -0.069&0.238 & 3.853 & -- \\
$^{36}$Ar & 301.96 & 306.06&306.95  &306.72&3.566&3.394&3.391 &0.021 &-0.255& -0.088 &0.256  &4.762 & -- \\
$^{38}$Ar & 326.59 &327.77&327.77 & 327.34&3.516&3.402&3.403 &0.013 &0.000& -0.041& 0.163 & 0.757 & -- \\
$^{40}$Ar & 342.87 &345.67&344.33 & 343.81&3.471&3.430& 3.427&0.013 &-0.031&-0.090 & 0.251 & 0.941 &-- \\
$^{42}$Ar & 357.01 &361.23&359.68  & 359.34& 3.453&3.435&3.435 &0.135 &-0.032& 0.121  & 0.275 & 2.332 &0.106 \\
$^{44}$Ar & 370.84 &374.41&373.76  & 373.73& 3.433&3.425&3.445 &-0.149 &-0.144& 0.114&0.240 & 2.888 &1.251 \\
$^{46}$Ar & 382.61 &386.25&386.46 & 386.93& 3.419&3.404&3.438 &0.049 &-0.135& -0.093 & 0.175 &4.320 &2.897 \\
$^{48}$Ar & 390.76 &394.33&395.39 & 395.76& 3.433&3.427& &0.082
&-0.205& -0.109 &  & 5.004 &1.890 \\
$^{50}$Ar & 397.87 &400.72&402.39 & 402.4&3.446&3.452& &0.069
&-0.240& -0.085 & & 4.530 &0.053 \\
$^{52}$Ar & 404.33 &406.06&406.41  & 406.59& 3.466&3.507& &0.065 &-0.281& -0.081  & & 2.262 &0.053 \\
$^{54}$Ar & 409.35 &408.99&408.88& & 3.503&3.554& &0.078
&-0.276& -0.130& & & 0.393 \\
$^{56}$Ar & 414.38 &411.68& 410.15&  &3.543&3.602& &0.064
&-0.212& -0.188 &  & & 0.766 \\
$^{58}$Ar & 419.67 &413.11& 410.32 & &3.583&3.641& &0&-0.222&  -0.191 & & & 1.282 \\
$^{60}$Ar & 418.26	& 413.57 & 409.88 & & 3.599 & 3.67 & & 0.109 & -0.253 & -0.197 & & & --\\
\hline
\end{tabular}
\label{tab1}
\end{table*}

\section{Results and Discussions}
\label{result}
The present work aims to establish the possible correlation between the nuclear structure and reaction in fusion studies. As explained in the preceding section, the RMF formalism can be used to successfully forecast the structural characteristics of finite nuclei close to the dripline \cite{maje22,yada02,yada04}. On the other hand, to calculate the reaction cross-sections, the basic ingredient in the Glauber model is the density, a structural observable. Therefore, carefully estimated charge and density distributions are necessary for precisely predicting the reaction cross-sections. Before proceeding to obtain the reaction characteristics of the systems associated with Ar-nuclei, the structural properties such as binding energy, root-mean-square charge radii, quadrupole deformation parameter, shell closure parameter, two neutron separation energy, differential two neutron separation energy and the depletion parameter are evaluated for $^{30-60}$Ar. It is worth mentioning that neutron-rich Ar-isotopes are predicted as nuclear halo \cite{kost19}, which will be of parallel interest to observe its relative effect in the reaction cross-section. The calculated results from RMF (NL3$^*$) are compared with the corresponding predictions from the finite range droplet model (FRDM) \cite{moll15}, Weizs$\ddot{a}$cker (WS) mass formula (WS3 and WS$^*$) parameters \cite{liu11,wang10} and the available experimental data \cite{wang12,ange13,rama01}.

\begin{figure}[H]
\centering
\includegraphics[width=0.8\columnwidth]{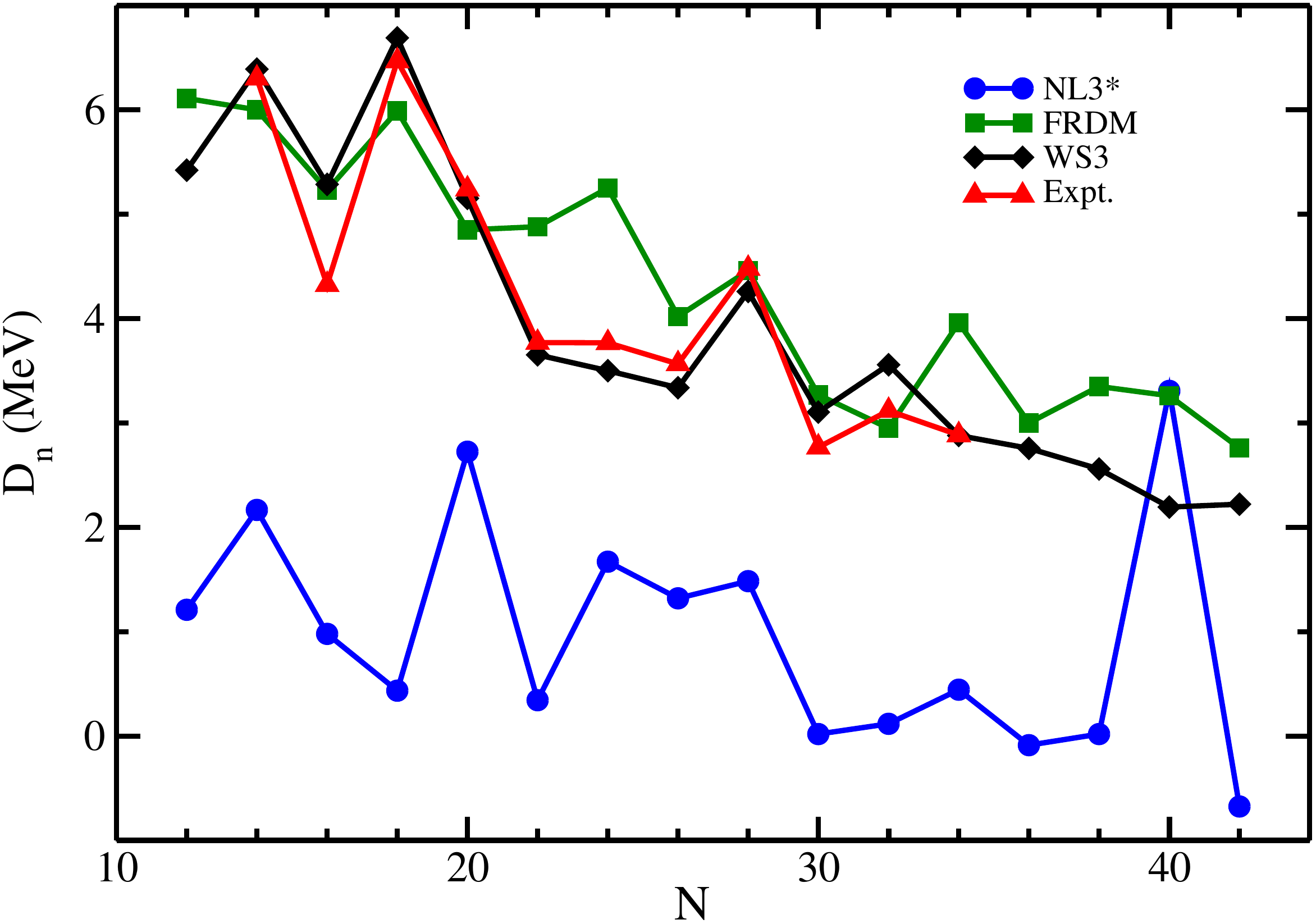}
\caption{\label{fig1}(colour online) The shell closures parameter (D$_n$) over the isotopic chain of $^{30-60}$Ar using relativistic mean-field for NL3$^*$ parameter set (Blue Circle) compared with FRDM predictions (Green Square), WS mass formula with WS3 parameter (Black Diamond) and available experimental data (Red Triangle) \cite{wang12}.}
\end{figure}
\subsection{Binding Energy and Nuclear Radius}
The binding energy is the amount of energy required to disassemble a nucleus into its constituent nucleons. As a result, it offers a detailed knowledge about the nuclear structure and serves as a major ingredient in obtaining the separation energies of neutron and proton as well as the shell closures of the participating Ar-isotopes. The calculated RMF (NL3$^*$) binding energies are compared with those from the FRDM \cite{moll15}, the Weizs$\ddot{a}$cker mass formula with WS3 parameter \cite{liu11} and the available experimental data \cite{wang12} as shown in Table \ref{tab1}. The second to fourth columns of the table reveals that the RMF(NL3$^*$) results reasonably agree with the other theoretical predictions and the available experimental data. From the table, one can observe a direct proportionality between BE and the mass number and/or neutron of the nuclei. In other words, the binding energy increases with an increase in mass number. However, a more detailed inspection shows a relatively higher magnitude in the binding energy per nucleon (BE/A)  for $^{38}$Ar-isotope over the isotopic chain.

In the same vein, the RMF (NL3$^*$) estimate of the charge radii (R$_{ch}$)  and quadrupole deformation ($\beta_{2}$)  for $^{30-60}$Ar isotopes are compared with the  WS$^*$, WS3 and the available experimental data \cite{ange13,rama01} as shown in the 6$^{th}$ to 12$^{th}$ columns of Table \ref{tab1}. Unlike the binding energies, the charge radius displays a peculiar behaviour i.e. the R$_{ch}$ values decrease with the increase of mass number up to A = 46, then it starts increasing with mass number. Correlating the quadrupole deformation parameter with nuclear radii, it is evident that there is a possibility of shape transition from prolate to oblate for $^{44}$Ar, and that causes the trends in the radius over the isotopic chain. Furthermore, the calculated results for  $\beta_{2}$ are found to be more consistent with the experimentally measured data than the FRDM and WS3 predictions.  Although the RMF (NL3$^*$) fairly reproduces the trend of R$_{ch}$ from the experimental measurement, there is a difference of 0.015 - 0.286 which is more pronounced at N= 14, 20, 40, unlike the theoretical predictions from the microscopic-macroscopic  WS$^*$ whose relative difference is between (0.006 - 0.034). Though the experimental data used here do not indicate N= 14, 20, 40 as possible sub-shell closures, the presence of sub-shell closures has been experimentally observed in \cite{craw22} and predicted theoretically in \cite{yada22,das22,patt22}. 

Here we have estimated the $\Delta BE = ({BE})_{Exp.} -  ({BE})_{Th.}$, which is defined as the difference between the binding energy of experimental data and our theoretical RMF (NL3$^*$) results and these values are tabulated in Table \ref{tab1}. The magnitude of $\Delta$BE increases from $^{30}$Ar to $^{36}$Ar, a sudden fall noticed at $^{38}$Ar, after this again it increases up to $^{48}$Ar. Here almost all the $\Delta$BE values are positive which means our RMF (NL3$^*$) results underestimate the experimental data.

\begin{figure}[H]
\centering
\includegraphics[width=0.8\columnwidth]{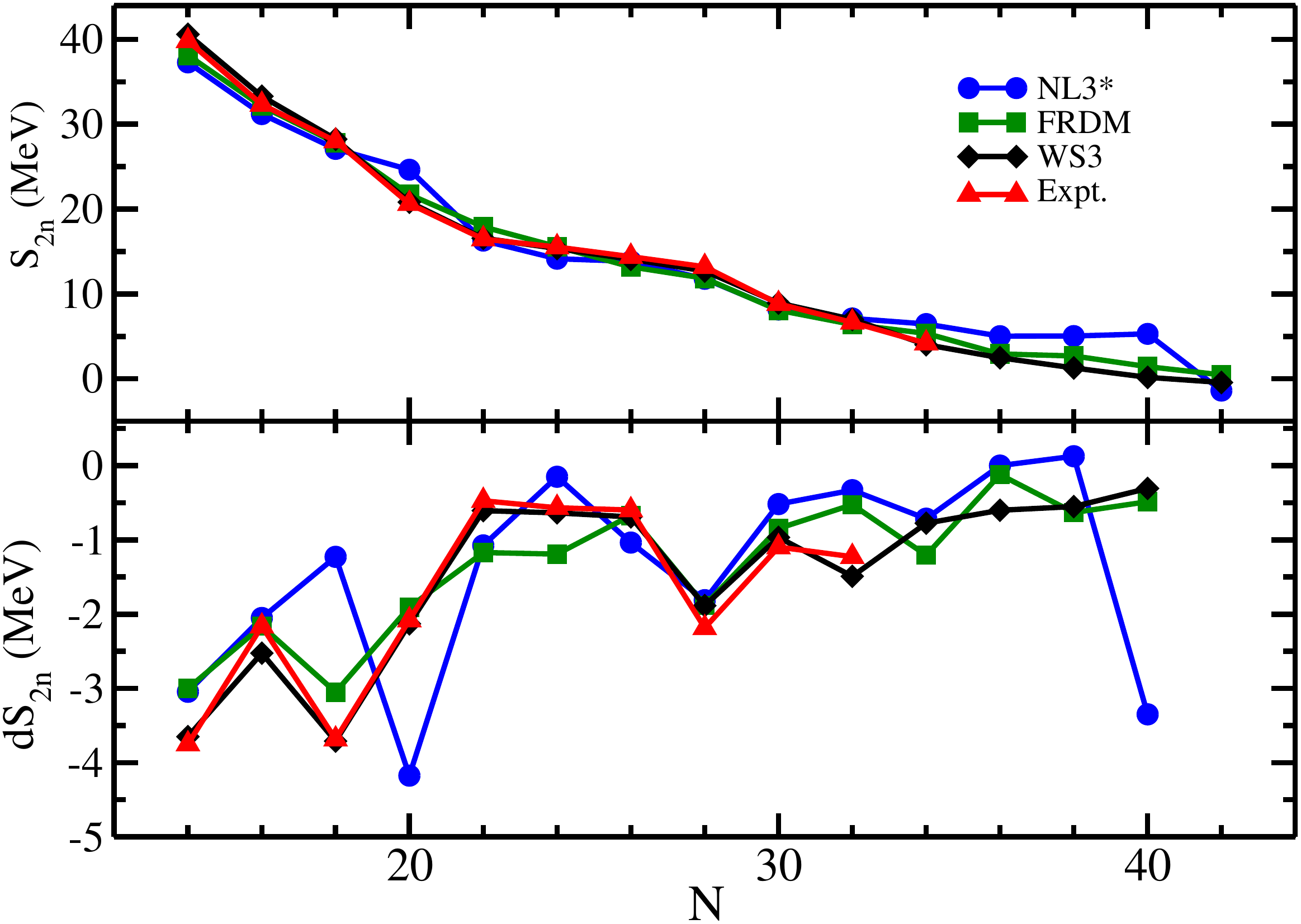}
\caption{\label{fig2}(colour online) The two neutron separation energy and differential two neutron separation energy as a function of neutron number for $^{30-60}$Ar using RMF (NL3$^*$) parameter set (Blue Circle) is compared with FRDM results (Green Square), WS3 mass formula (Black Diamond) and available experimental data \cite{wang12} (Red Triangle).}
\end{figure}

\subsection{Shell Closure Parameter and Separation Energy}
The indication of shell closure or its collapse can be investigated using the recently introduced physical quantity D$_n$(N) given by the formula \cite{brow13,thak19}
\begin{eqnarray}
\label{shell}
D_{n}(N) &=& (-1)^{N+1} [S_{n}(Z,N+1)- S_{n}(Z,N)] \nonumber \\
&=& (-1)^{N} [2BE(Z,N) - BE(Z,N-1)  - BE(Z,N+1)].
\end{eqnarray}
Here, the involved proton (Z), neutron (N), and mass number (A) of a nucleus are used to calculate the single neutron separation energy S$_{n}$(Z, N) = BE(Z, N) - BE(Z, N - 1). The variation of shell closure parameter D$_{n}$(N) with neutron number is plotted in Fig. \ref{fig1} for $^{30-60}$Ar-isotopes. The estimated values for RMF (NL3$^*$) (blue line and solid circle) parameter set are compared with the predictions from the FRDM (green line and solid squares), WS3 (black lines and solid diamond) as well as the available experimental data (red line with solid triangle). Notable peaks are spotted at N = 14, 20, and 40 for RMF (NL3$^*$) signifying the presence of shell closures. At certain points, one can observe the discrepancy in the trends between calculated and experimental curves. In the present study, we are unable to reveal the exact observable that causes the deflections, which is beyond our present study. The appearance of shell/sub-shell gap at $Z$ or $N=14$ is traceable to the filled $0d_{5/2}$ orbit in $^{32}$Ar nucleus in the ground state. On the other hand, small peaks at N = 24, 28, and 34 indicate weak shell closures within the Ar-isotopic chain. These results agree well with the theoretical predictions of Refs. \cite{adri20,thak20}.  Similarly, the FRDM and WS3 predictions are characterized by a downward kink at N=20 and conspicuous peaks are found at N = 18, 24, 28, 34, 40 and N = 14, 18, 28, 32 respectively.

Furthermore, we calculated the traditional observable for shell/sub-shell closure over an isotopic chain, i. e.,  two neutron separation energy. It delineates the amount of energy required to remove two neutrons from a nucleus and is given in a simple relation, $S_{2n} = BE (Z, N) - BE (Z, N-2)$. The S$_{2n}$ is calculated by using the binding energies from the respective models and available experimental data and is shown in Fig. \ref{fig2}. From the figure, the calculated S$_{2n}$ values decrease with an increase in neutron number along the isotopic chain except for an enhanced kink and a significant drop of about 2-6 MeV at $^{32}$Ar, $^{38}$Ar and $^{58}$Ar. This again confirms the indication of shell/sub-shell closure at $N=14,20$ and $40$, which appeared in $D_n$ (see Fig. \ref{fig1}). The same is true for $^{46}$Ar $(N=28)$ although it manifests a disappearing/eroding kink, alluding to its weak shell closure. Besides the slight variations at these nodes, the RMF(NL3$^*$) results are consistent with the FRDM and WS3 predictions, as well as the experimental data and the previous studies of Ref. \cite{adri20}. 
Further, the differential two neutron separation energy formula is estimated by using the relation, $dS_{2n} = \frac{S_{2n}(Z, N+2)-S_{2n}(Z, N)}{2}$. It provides the relative change and/or rate of change in the separation energy with regard to the isotopic chain's neutron count. Additionally, information regarding the presence or collapse of a shell closure enriched it. The calculated results of dS$_{2n}$ using the RMF(NL3$^*$) are found to be consistent with those from the FRDM and WS3 with the available experimental data shown in Fig. \ref{fig2}.  One obvious observation is that the RMF(NL3$^*$) predictions seem to uniquely identify the traditional magic number N = 20 among others. As earlier mentioned the results of the  RMF(NL3$^*$) adapt well with recent theoretical and experimental findings \cite{yada22,das22,patt22}. 
Recent findings \cite{craw22} give new insight into the presence of a sub-shell closure at $N=14$. Generally, a small deviation in $S_{2n}$ in Fig.\ref{fig2} yields a large difference in its derivative  $dS_{2n}$ (which can be understood from a mathematical viewpoint). 

\begin{figure}[H]
\centering
\includegraphics[width=0.9\columnwidth]{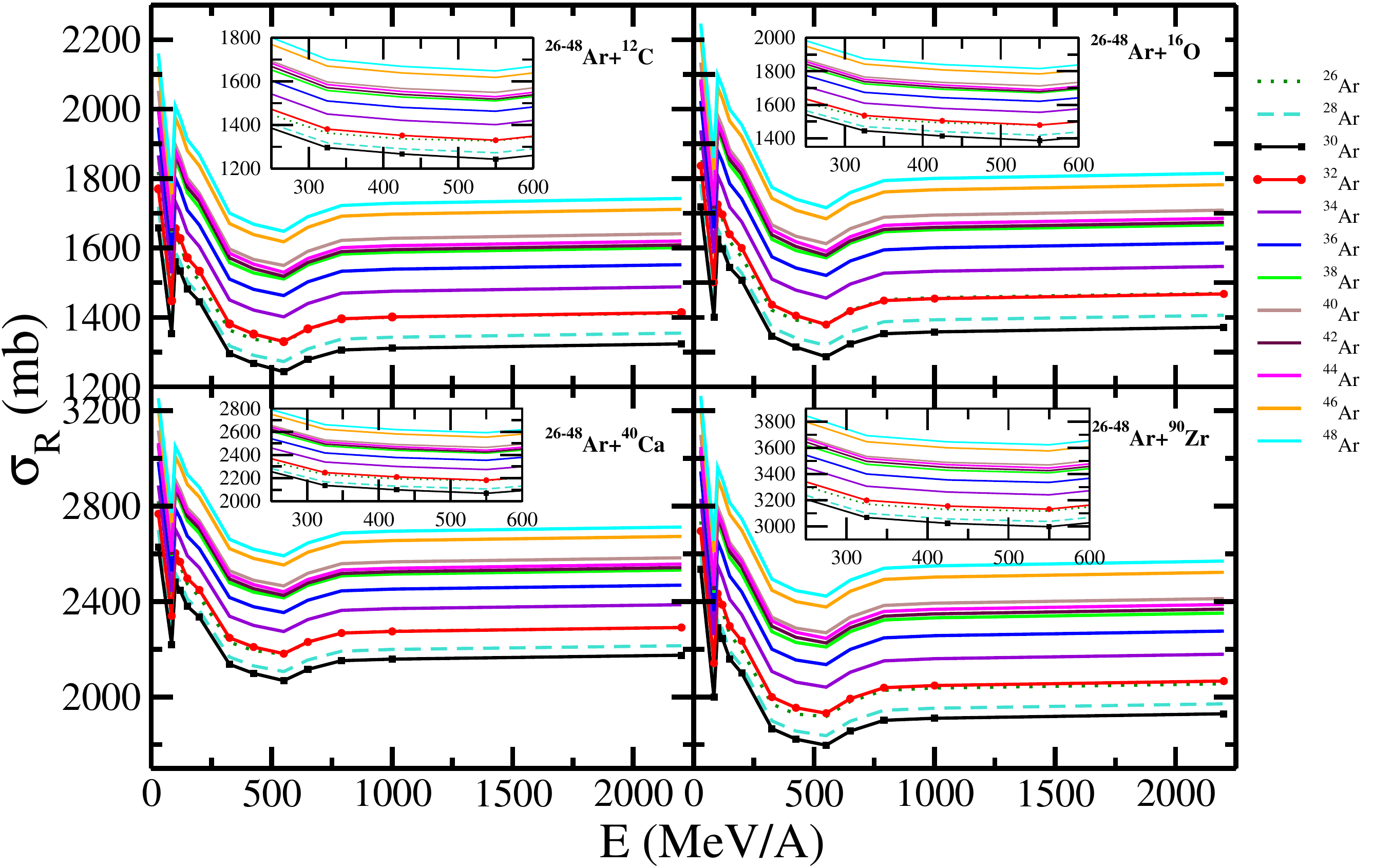}
\caption{\label{fig3}(color online) The reaction cross-section for $^{26-48}$Ar as projectiles with targets as $^{12}$C, $^{16}$O, $^{40}$Ca, $^{90}$Zr for different projectile energies.}
\end{figure}

\subsection{Reaction Cross-Section} 
\label{cross-sect}
To assess the total reaction ($\sigma_{R}$) and differential cross-section $(d\sigma /d\Omega)$ for the combination of an unstable projectile with a stable target and/or a stable projectile with an unstable target remains an unresolved problem in nuclear physics. Furthermore, a detailed analysis of the $\sigma_{R}$ and $(d\sigma /d\Omega)$ provides a rich comprehension of the nuclear structure and a valuable insight into the evolution of drip-line nuclei. At present, various studies \cite{shuk07,shar06,patr09} have alluded to the successful applicability of the RMF densities within the Glauber model in terms of its ability to reproduce the experimental cross-section for different regions of the nuclear chart. Here we have taken the light mass isotope as the projectile and both light as well as heavy mass isotopes as the target to study the reaction characteristics. Further, the study aims to correlate the structural effect such as shell/sub-shell closure and nuclear halo effect in the reaction dynamics.

\begin{figure}[H]
\centering
\includegraphics[width=0.9\columnwidth]{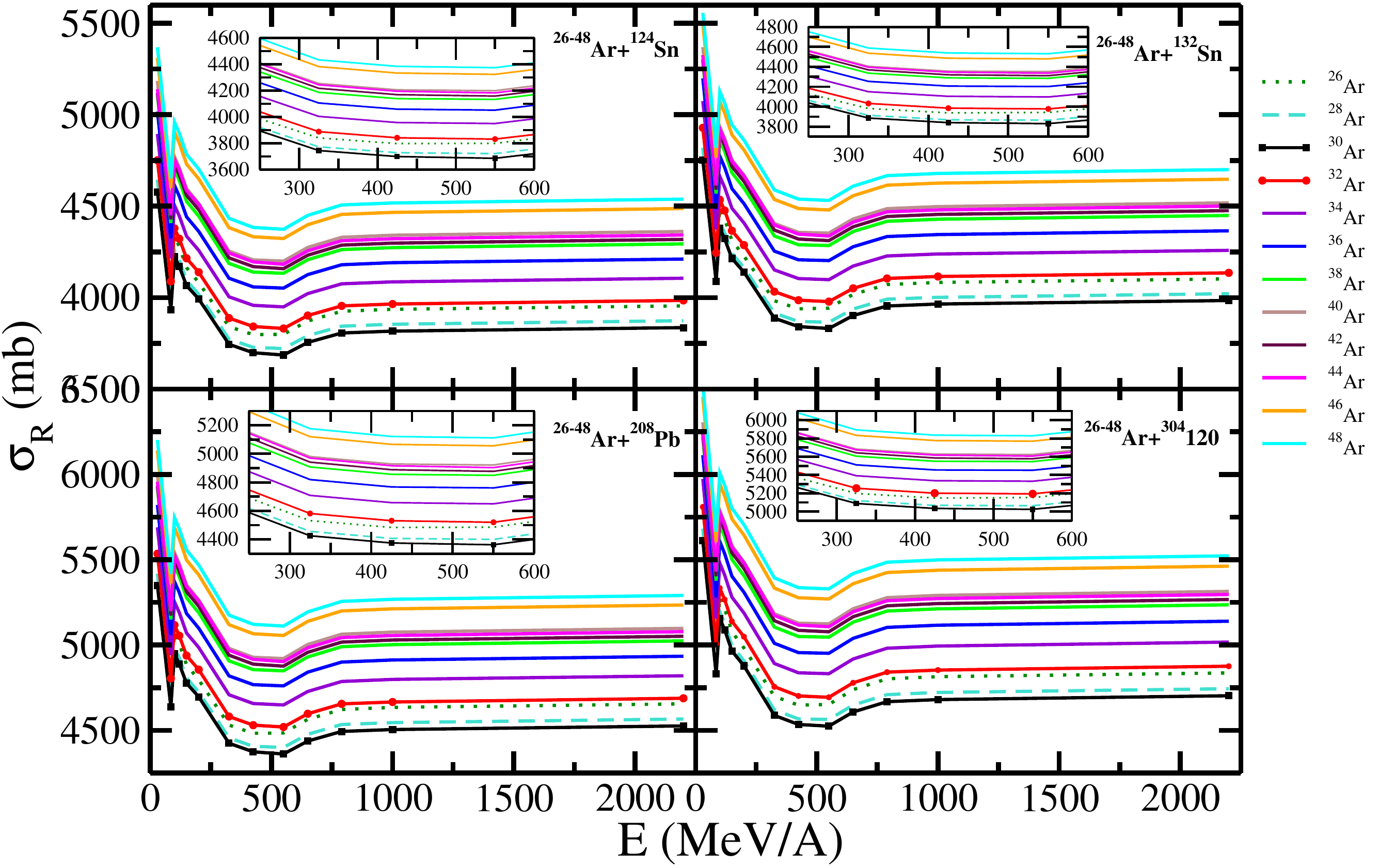}
\caption{\label{fig4}(color online) The reaction cross-section for $^{26-48}$Ar as projectile with target as $^{124}$Sn, $^{132}$Sn, $^{208}$Pb, $^{304}$120 at different projectile energy.}
\end{figure}

Fig. \ref{fig3} shows the profile of the total reaction cross-section for Ar-isotopes ($^{26-48}$Ar) as projectiles and $^{12}$C, $^{16}$O, $^{40}$Ca, $^{90}$Zr as targets at different incident energies. From the figure, it is clear that the total reaction cross-section decreases to 550 MeV/A of incident energy and is accompanied by a gentle rise to about 760 MeV/A and after that, it assumes a nearly constant value. A relative higher cross-section is found in between $^{30}$Ar and $^{32}$Ar-isotopes for all the targets ($^{12}$C, $^{16}$O, $^{40}$Ca and $^{90}$Zr) taken in the Fig. \ref{fig3}. It may be an indication that $^{32}$Ar is a close shell/sub-shell isotope. By comparing, the four combinations in the figure infer that $\sigma_{R}$ increases as the mass of the projectile increases. In each case, the inset shows the magnified view of the total reaction cross-section between 200-600 MeV. The total reaction cross-section for $^{26-48}$Ar as projectiles with $^{124,132}$Sn, $^{208}$Pb, $^{304}$120 as targets is shown in Fig. \ref{fig4}. It is worth noting that, the difference between Fig \ref{fig3} and Fig. \ref{fig4} is the consideration of light and heavy targets, respectively for the same projectiles. Besides the similarity in the profiles in which an abrupt decline is noticed in the total reaction cross-section at 550 MeV/A and its attending trend, a wider cross-section gap value is observed between $^{30}$Ar and $^{32}$Ar which strengthens the opinion that $^{32}$Ar to be a magic isotope in nature. In other words, $\sigma_{R}$ is found to be maximum at low incident energies of the projectile nucleus and undergoes a sharp drop as the energy increases forming minima at 550 MeV/A. 

Further increases in the incident energy may initiate a rise in the total reaction cross-section followed by saturation. Apart from this, it is well known  that larger values of the reaction cross-section can give a clearer insight into the halo characteristics of the considered isotopes. Here, we have calculated the $\sigma_{R}$ values for a series of Ar-isotopes as projectiles with different light to heavy mass targets and we are seeing that the $\sigma_{R}$ values increase with the increase of projectile and target mass and also a relative higher cross-section values in between $^{30}$Ar - $^{32}$Ar isotopes may give a concluding remark regarding the halo nature of these isotopes for all the targets. We reiterate that Fig. \ref{fig3} and \ref{fig4} are the variations of the total reaction cross-section ($\sigma_{R}$) to incident energy. At lower incident energies the decreases with incident energy and with an increase of the energy the reaction cross-section values attain their maximum. The $\sigma_{R}$ values decrease with energy up to around 100 MeV/A then increase up to 250 MeV/A. After this it attains stability. Although currently unsynthesized,  $^{304}$120 is a predicted stable nucleus from various theoretical studies e.g.  Ref. \cite{ahma12,bhu12}. 

\begin{figure}[H]
\centering
\includegraphics[width=0.9\columnwidth]{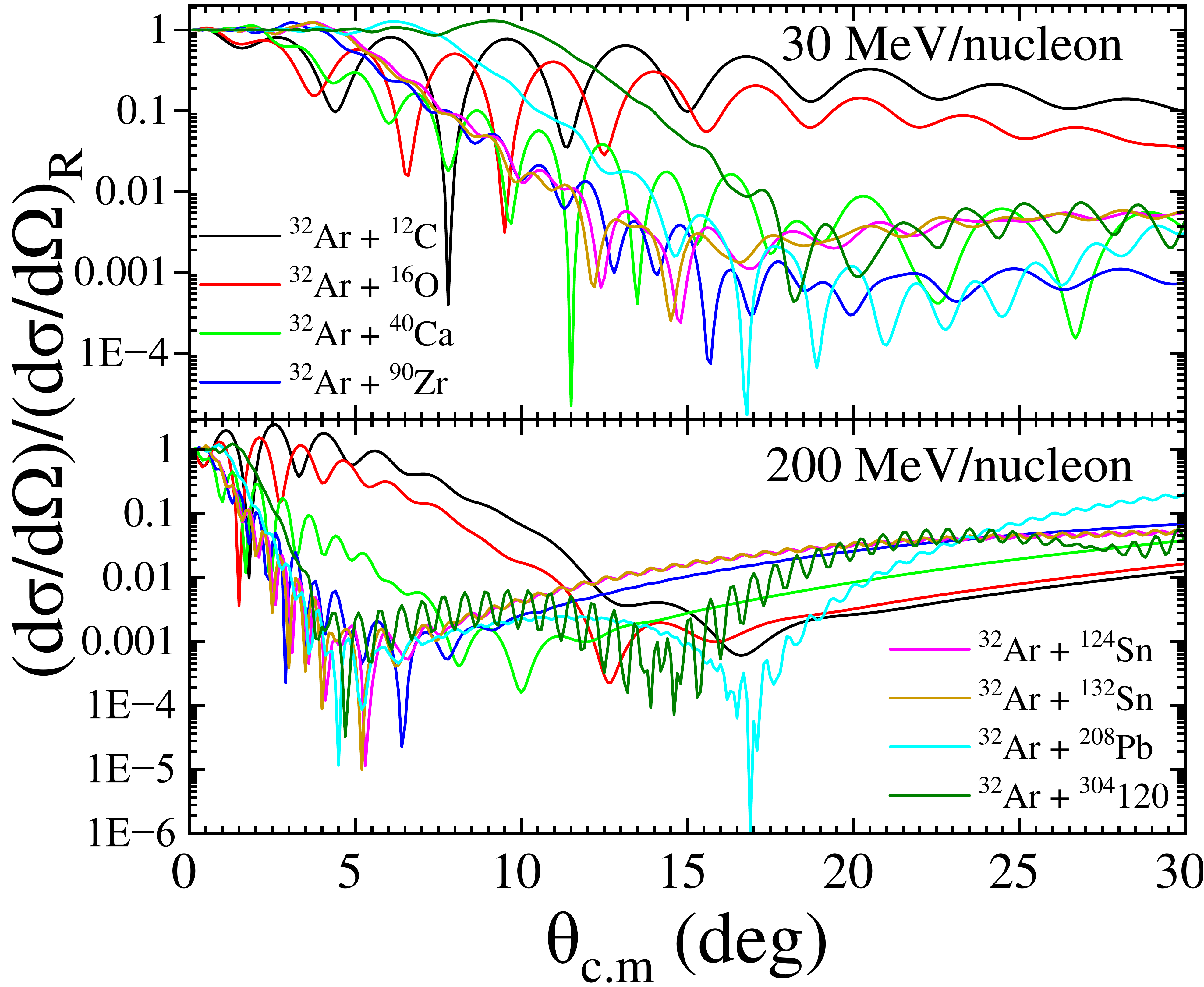}
\caption{\label{figa} (Color online) The differential cross-section for $^{32}$Ar as the projectile and $^{12}$C, $^{16}$O, $^{40}$Ca and $^{90}$Zr, $^{124,132}$Sn, $^{208}$Pb and $^{304}$120 as targets at incident energies 30, 200 MeV/A with respect to scattering angles are framed in the upper and lower panel respectively.}
\end{figure}

\begin{figure}[H]
\centering
\includegraphics[width=0.9\columnwidth]{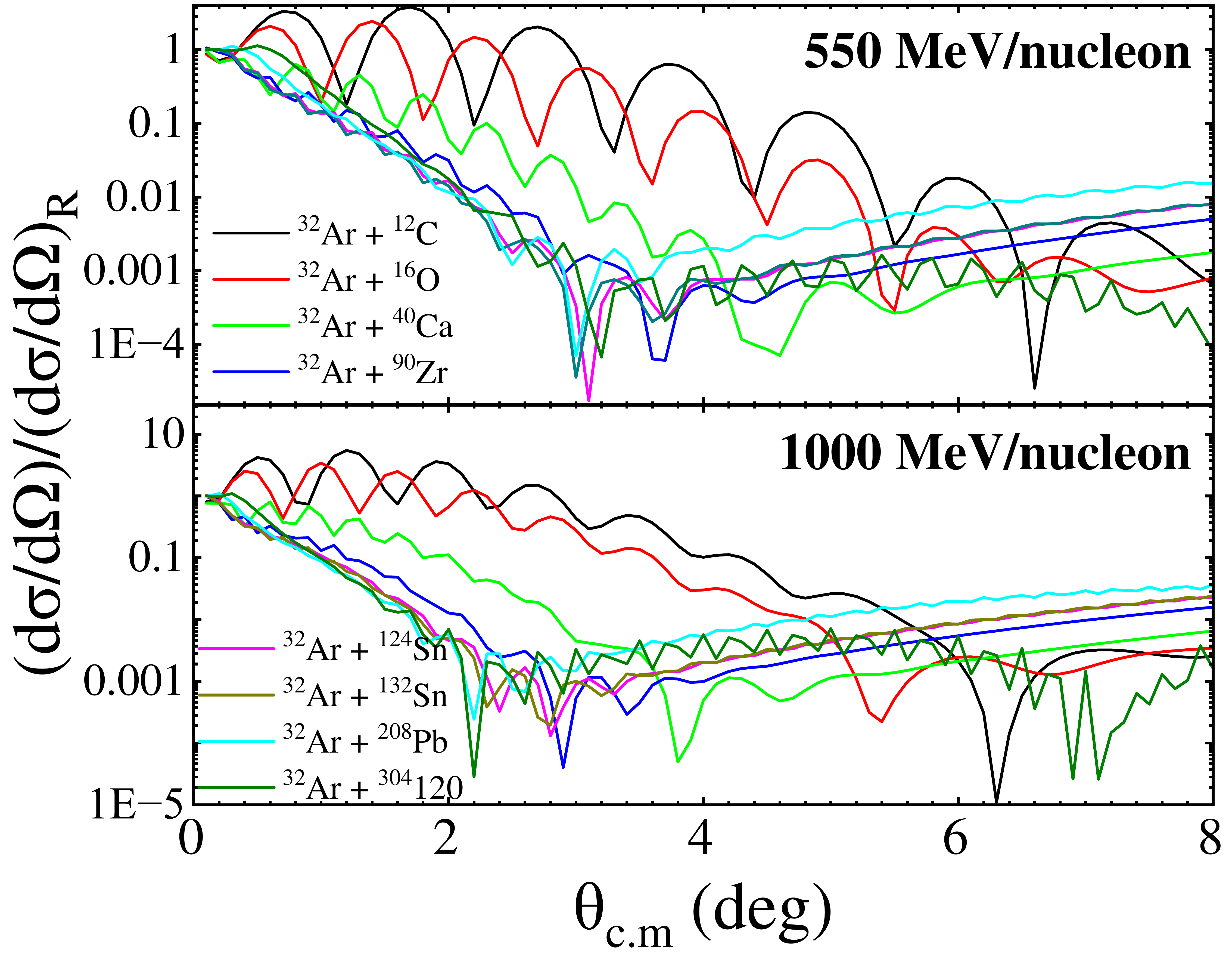}
\caption{\label{figb} (colour online) Same as Fig. \ref{figa} but the incident energies are 550 and 1000 MeV/A. 
}
\end{figure}

The differential cross-section $(d\sigma /d\Omega)$ is shown in Eq. (\ref{diff}), another essential observable in the reaction study that unfolds the scattering phenomena. Categorically, the first three targets $^{12}$C, $^{16}$O and $^{40}$Ca are N=Z nuclei and other targets such as $^{90}$Zr (Z=40, N=50), $^{124}$Sn (Z=50, N=74), $^{132}$Sn (Z=50, N=82), $^{208}$Pb (Z=82, N=126), $^{304}$120 (Z=120, N=184) are neutron-rich with $^{32}$Ar-isotope as the only projectile are implemented to calculate the differential reaction cross-section values. It is worth noting that in $N=Z$ nuclei, one would anticipate that neutrons and protons would work together to deform the nucleus towards its most confined configuration \cite{list87,patr02}. Some of these orbits have noticeable gaps between them, which suggests the presence of high-binding energy shells. The so-called magic number in such a shell is the greatest number of nucleons that can fill orbits from the first subshell to the appropriate one \cite{laou20}. The $d\sigma /d\Omega$ for this isotope is examined at various incident energies likely 30, 200, 550, 1000 MeV/A using RMF (NL3$^*$) approach in the upper and lower panel of Fig. \ref{figa} and Fig. \ref{figb} respectively. Figs. \ref{figa} and \ref{figb} depict the differential cross-section profile for target nuclei $^{12}$C, $^{16}$O, $^{40}$Ca, $^{90}$Zr, $^{124,132}$Sn, $^{208}$Pb and $^{304}$120 as target nuclei and the $^{32}$Ar as a projectile for all the considered incident energies (30, 200, 550, 1000 MeV/A). For each of the mentioned incident energies, a deep deposition and Fresnel diffraction pattern is noticed around 0-10$^{\circ}$ arising from the interference of the Coulomb and nuclear amplitudes. These damping oscillations are repeatedly accompanied by an exponential increase in differential cross-section value for both $^{12}$C and $^{16}$O targets with all the incident energies except at 30 MeV/A. The oscillatory nature is detected for $^{12}$C and $^{16}$O targets for 30 MeV/A energy beyond the scattering angle of 10$^{\circ}$ as shown in Fig. \ref{figa}.

The pattern of oscillation in $d\sigma /d\Omega$ at high scattering angles substantiates the findings of \cite{patr09}. In Fig. \ref{figa} and Fig. \ref{figb}, at lower energy, the differential cross-section for $^{40}$Ca and $^{90}$Zr targets manifest Fresnel diffraction pattern and deep deposition  at $\theta_{c.m}$ = 10-15$^{\circ}$. Specifically, for $^{32}$Ar + $^{40}$Ca reaction at higher energies, a deep deposition is noticed at scattering angles 4-4.5$^{\circ}$, 3-4$^{\circ}$ for 550 and 1000 MeV/A respectively. For the reaction, $^{32}$Ar + $^{90}$Zr, the respective energies and scattering angles are  200, 550, 1000 MeV/A, 6$^{\circ}$, 3-4$^{\circ}$ and 2-3$^{\circ}$ where deep deposition occurs and after that, a sharp increase in the differential scattering cross-section values is noticed at higher energies in case of $^{40}$Ca and $^{90}$Zr targets. Similarly, the Fresnel diffraction for the reaction $^{32}$Ar + $^{124,132}$Sn at different incident energies shown in the upper and lower panel of Fig. \ref{figa} and Fig. \ref{figb}. Furthermore, deep position is spotted at scattering angle 0-10$^{\circ}$ for $^{32}$Ar + $^{124,132}$Sn with incident energies 200, 550 and 1000 MeV/A. For the larger value of the scattering angle, a sharp increase in differential cross-section is observed.

Besides, in Fig. \ref{figa} and \ref{figb}, for $^{208}$Pb target two deep depositions pointed at $\theta_{c.m}$ = 0-5 $^{\circ}$ and 10-15 $^{\circ}$ for 200 MeV/A scattering energy. At incident energy 550 MeV/A and 1000 MeV/A, the deep is formed between 0-5 degrees. As previously observed, the oscillatory nature of differential cross-section is seen at 30 MeV/A between 10-15$^{\circ}$. Among all the differential cross-section values using various targets, it is noted that for $^{208}$Pb case, the longest-tailed d$\sigma$/d$\Omega$ values are for 30 and 200 MeV/nucleon. Whereas for 550 and 1000 Mev/nucleon energies, a long tail differential cross-section has marked for $^{12}$C as targets. Moreover, vigorous oscillations (maxima-minima) are perceived for higher mass targets i.e $^{304}$120 within the energies 200, 550, 1000 MeV/A shown in Fig. \ref{figa} and Fig. \ref{figb}. The Fresnel type of diffraction is also noticed at 0-10$^{\circ}$ in the case of $^{304}$120 target. Thus, the nature of the target nuclei and the magnitude of the incident energy decides the profile of the differential cross-section and the scattering angle. 

\begin{figure*}
\centering
\includegraphics[width=1.1\columnwidth]{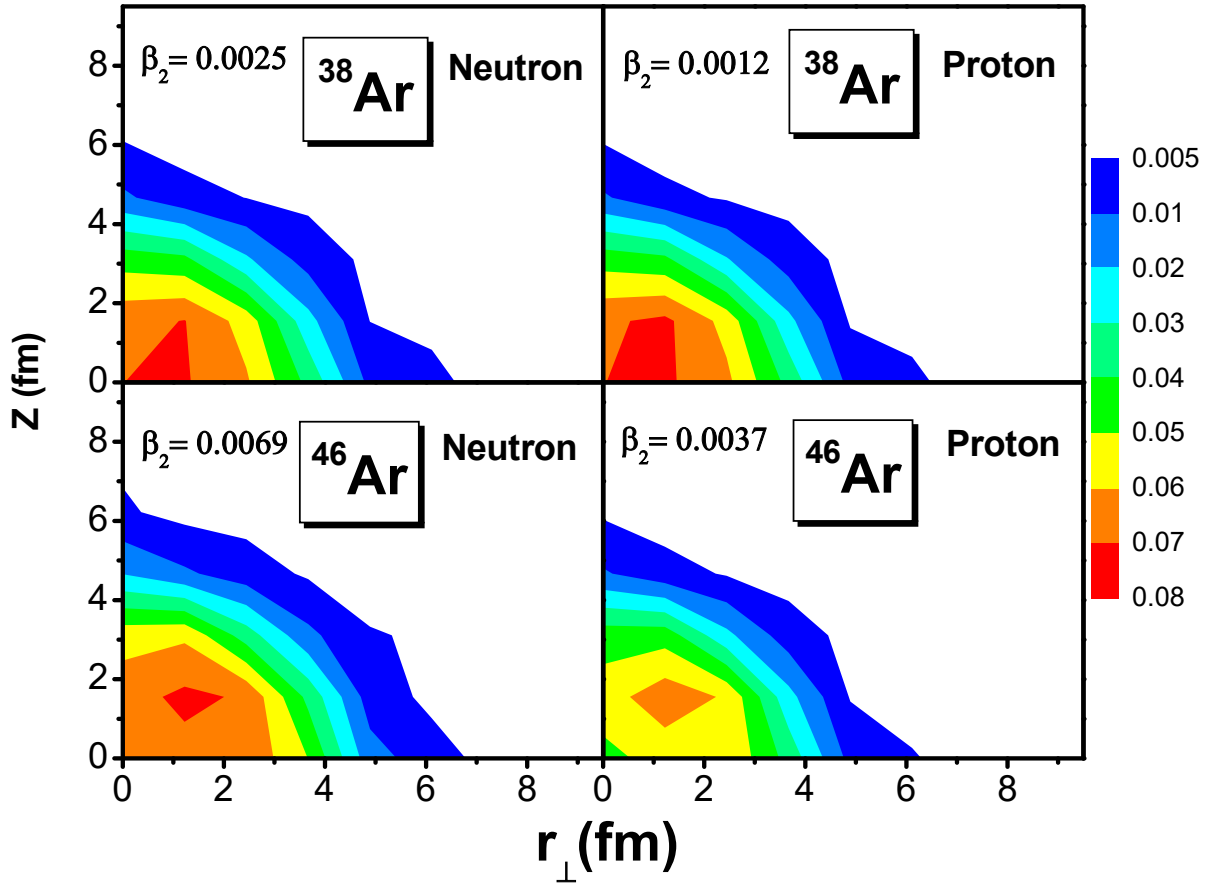}
\caption{\label{figx}(color online) The profile of the proton and neutron density distribution for even-even $^{38,46}$Ar isotopes as representative cases. The dark blue colour corresponds to the lowest density ($ \rho\approx0$) while the red depicts the maximum density as indicated in the colour bar.  A similar pattern was observed in $^{48-60}$Ar-isotopes which are not shown here for the sake of brevity. 
}
\end{figure*}

\subsection{Depletion Fraction}
The appearance of a bubble or semi-bubble structure can be quantified by the depletion fraction (D.F) \cite{gras09,gras09a} which is expressed in percentage (\%)  as,
\begin{eqnarray}
(D.F)_T = \frac{(\rho_{max})_T-(\rho_{cent})_T}{(\rho_{max})_T}*100.
\label{dfeqn}
\end{eqnarray}
Here, $\rho_{max}$ and $\rho_{cent}=\rho(r=0)$ are the maximum and central neutron density and $T$ stands for total (= neutron+proton) density distribution \cite{co18,shuk14,shuk11,shar15}. The RMF formalism is employed here to account for the bubble formation in the chosen systems. The total D.F for Ar-isotopes is tabulated in Tab. \ref{tab1}. The blank spaces in the table correspond to the Ar-isotopes with $D.F = 0$ means not having the bubble like structure, and  $D.F > 0$ signify the existence of semi-bubble nuclei. It is seen from the Tab. \ref{tab1} that $^{32}$Ar and $^{42-58}$Ar-isotopes have $>$ 0 total D.F values, indicating them to be semi-bubble nuclei. Among these isotopes of Argon, $^{46}$Ar has the highest total depletion fraction value. For a detailed analysis, the depletion fraction in deformed nuclei could be explained in terms of their symmetric $(z)$ axis \cite{kuma22}. The profile of the total radial density distribution and/or internal configuration of $^{38,46,50,60}$Ar isotopes are presented in Fig. \ref{figx} along their $z$-axis as representative cases.  The same trend as of Fig. \ref{figx} can be obtained for other examined isotopes. The dark blue colour corresponds to the lowest density ($\rho\approx0$) while dark red depicts the maximum density. In all cases, the central density depletion is obvious as $ \rho$ tends to be approximately zero. Our results further reveal that nuclear deformation plays a significant role in bubble formation. Particularly, we found that large deformations elicit an abrupt decline in the depletion fraction. Other than  $^{38}$Ar $(N=20)$ and $^{46}$Ar $(N=28)$, a careful inspection shows that the depletion fraction reduces as deformation increases owing to the less conspicuous shell effect. This is in line with the recent findings of  Kumar {\it et al.} \cite{kuma22}, which showed that the valence nucleons control structural change brought on by deformation. In principle, the decrease in the depletion fraction with increasing neutron number can be attributed to the influence of isospin and the co-action of the Coulomb and nucleon-nucleon interaction on nuclear bubbles. 


\section{Summary and Conclusions}
\label{summary} 
We have investigated the correlation between the nuclear structure and reaction in fusion studies using the Ar-isotopes within the mass range  30 {$\leqslant$} A {$\leqslant$} 60 to probe the nuclear structure. We employ the relativistic mean-field (RMF) with the NL3$^*$ parameter set. As such, the bulk properties such as the binding energy (BE), charge radii (R$_{ch}$), quadrupole deformation parameter ($\beta_2$), shell closure parameter, two neutron separation energies ($S_{2n}$), differential two neutron separation energies ($dS_{2n}$) are analysed and the results are compared with the predictions from FRDM, WS formula with WS3, WS$^*$ parameter and the available experimental data. From the comparison, it can be shown that the RMF (NL3$^*$) yield a reasonable agreement with both the theoretical and experimental data. From the structural analysis, most of the considered isotopes are prolate shape and $^{38}$Ar is the most stable isotope. The relative difference between the RMF binding energies and the experimental BE i.e the $\Delta$BE further suggests that our results are consistent with the experimental data. The shell closure parameter (D$_{n}$(N)) and the two neutron separation (S$_{2n}$) energy shows that A = 32, 38, 58 are shell closure nuclei and A = 42, 46, 52 are weak shell closure nuclei.

The estimated total depletion fraction suggests that $^{32}$Ar and $^{42-58}$Ar-isotopes are semi-bubble nuclei. However, using the RMF densities for the neutron-rich light mass nuclei and the $^{26-48}$Ar isotopes as projectiles with stable targets $^{12}$C, $^{16}$O, $^{40}$Ca, $^{90}$Zr, $^{124,132}$Sn, $^{208}$Pb, and $^{304}$120 as for various projectile energies, the Glauber model is used to estimate the reaction cross-section and differential scattering cross-section. The differential scattering cross-section $(d\sigma /d\Omega)$ for a combination of $^{32}$Ar-isotope and all the stable targets are also analysed. As such, we conclude that the scattering angle ($\theta_{c.m}$) and  $(d\sigma /d\Omega)$ strongly depend on the mass of the target nuclei as well as the magnitude of the incident energy of the projectile. This unequivocally builds a bridge between the nuclear structure and nuclear reaction.  

\end{document}